\documentclass[structabstract]{aa}  
\usepackage{graphicx}
\usepackage{txfonts}
\usepackage{natbib}
\bibpunct{(}{)}{;}{a}{}{,}
\begin{document}
   \title{Spectroscopic study of early-type multiple stellar systems}
   \subtitle{II. New binary subsystems}

   \author{M. E. Veramendi \and J. F. Gonz\'alez 
          }

   \institute{Instituto de Ciencias Astron\'omicas, de la Tierra y del Espacio (CONICET-UNSJ), Casilla de Correo 49, 5400 San Juan, Argentina \\
              \email{mveramendi@icate-conicet.gob.ar}\\
              \email{fgonzalez@icate-conicet.gob.ar}
             }

   \date{Received ...; accepted ...}

\titlerunning{Early-type multiple stellar systems. II.}
\authorrunning{M. E. Veramendi and J. F. Gonz\'alez}

 
  \abstract
 { This work is part of a long-term spectroscopic study
   of a sample of 30  multiple stars with early-type components.
   In this second paper we present the results of six multiple systems in which new stellar
   components have been detected.} 
   {The main aim is to increase the knowledge of stellar properties and dynamical structure of early-type multiple stellar systems.
   }
   {Using spectroscopic observations taken over a time baseline of more than 5 years we 
   measured RVs by cross-correlations and applied a spectral disentangling method to double-lined systems. Besides the discovery of objects with double-lined spectra, 
   the existence of new spectroscopic subsystems have been inferred from the radial  
   velocity variations of single-lined components and through the variation of the
   barycentric velocity of double-lined subsystems.
   Orbital elements have been calculated when possible. }
   {Seven new stellar components and two members that we expect to confirm with new observations 
   have been discovered in the six studied multiples.
   We present orbital parameters for two double-lined binaries and preliminary orbits
   for three single-lined spectroscopic binaries.
   Five of the six analysed systems are quadruples, while the remaining has five components
   distributed in four hierarchical levels. 
   These multiplicity orders are in fact lower limits, since these systems lack
   high-resolution visual observations and additional hierarchical level might exist in 
   that separation range.
   }
   { The six analysed systems have greater multiplicity degree and a more complex hierarchical structure than previously known, which suggests that high-order multiple systems are significantly more frequent that it is currently estimated.  
The long term spectroscopic monitoring of multiple systems has shown to be 
   useful for the detection of companions in intermediate hierarchical levels.}


   \maketitle
%

\section{Introduction}

The observed properties of multiple stellar systems (three or more components) are empirical evidence particularly useful to contrast theoretical models of stellar formation and evolution. However, the observational results published in literature on systems with early-type components are insufficient, both in quantity and detail, to establish overall conclusions on the frequency and properties of this type of systems \citep[see][hereafter Paper I, and references therein]{2014A&A...563A.138V}. For that reason, 
we carried out a spectroscopic investigation of a sample of 30 systems with catalogued components with types O, B, and A, selected from the Multiple Star Catalogue \citep[][hereafter MSC]{1997A&AS..124...75T}. In Paper I we explained in detail the selection criteria and  presented the full list of the analysed systems together with the information given in the MSC. In that paper we described the analysis and results obtained for six multiples for which we have determined orbital parameters for the first time or have corrected their previously published orbits. In this second paper we present the study and the results obtained for six systems in which we detected new components. 

In Sect.~2 we briefly outline the spectroscopic observations and radial velocity (RV) measurements, which were described in detail in Paper I. The analysis and results for the six systems studied in the present paper are presented in Sect.~3. In Sect.~4 we discuss the contribution of this work to the knowledge of the hierarchical structure of the studied systems. We summarize our main conclusions in Sect.~5.

\section{Observations and data analysis}

The spectroscopic observations were carried out using the 2.15 m telescope and the REOSC echelle spectrograph at Complejo Astron\'omico El Leoncito (CASLEO), San Juan, Argentina. In addition to the ten observing runs mentioned in Paper I, some systems presented in this paper were recently observed during three nights in July 2013. The spectra cover the wavelength range 3700--6300~\AA~with a resolving power $\mathrm{R}$=13\,300, and were reduced by using standard data reduction procedures within the NOAO/IRAF package.

Spectral types were determined by comparison with standard stars observed at the same spectral resolution. The procedure for measuring RVs has been described in Paper I.
Basically, we used cross-correlations and, in the case of double-lined spectroscopic binaries, the spectral disentangling method by \citet{2006A&A...448..283G}.

Based on the spectral morphology and the analysis of measured RVs, we evaluated the multiplicity of the observed stellar components. More specifically, a component with a single-lined spectrum was classified as RV variable if the rms deviation of the measured velocities was at least twice as large as the mean error of the measurements. When it was possible to determine the period of the variation and compute a preliminary orbit, the object was classified as single-lined spectroscopic binary (SB1). The observed components with rms deviation lower than twice the mean measurement error were considered RV constants on the time baseline of our observations. Finally, components whose spectra showed lines belonging to two stars with RVs linearly correlated with each other (consistent with orbital motion) were classified as double-lined spectroscopic binaries (SB2).

In new spectroscopic binaries we determined a preliminary orbital period using the {\it Phase Dispersion Minimization} technique \citep{1978ApJ...224..953S} as implemented in
 the IRAF task $pdm$,  which is suitable for irregularly spaced data. 
 We then applied the least-squares method to determine the elements of a Keplerian orbit, including recalculation of the period. 
Measurement errors were used to weight the data points. The parameter errors were calculated from the rms of the residuals.
In general, we fitted orbital period ($P$), time of primary conjunction ($T_0$), eccentricity ($e$), longitude of periastron ($\omega$), semi-amplitude of RV curve ($K$), and barycentre's velocity ($V_\gamma$). 

\section{Results}

\begin{table*}
\caption{Components of multiple systems observed and analysed in this paper.}
\label{table:1}
\centering
\begin{tabular}{c c c c c c} 
\hline 
\hline
System & Comp. & HD & ST$_{\rm pub}$ & ST & Result \\ 
\hline 
\noalign{\smallskip}
08079--6837   & Aa+Ab   & 68\,520 (AB) & B6IV        & B6\,IV+B8         & Orbit SB2 \\     
              & Ba+Bb   &            & 0.06        & A2\,V+A2\,V         & New SB2 \\          
\\
08314--3904   & Aa      & 72\,436 (AB) & B4V         & B4\,V             & New SB1 with preliminary orbit \\     
              & B       &            & -0.13       & B6\,V             & New RV variable \\ 
              & C       &            &             & F0\,V             & RV constant, probably non-member \\      
\\
10209--5603   & A       & 89\,890      & B3IIIe      & B3\,III         & RV constant (line profile variable) \\    
              & Ba+Bb   &            & B9IVpSi     & A0\,IVpSi+A2      & New SB2 (barycentric RV variation?)\\ 
              & C       &            &             & K0\,III           & RV constant \\ 
\\
13226--6059   & AB      & 116\,087     & B3V+...     & B3\,V             & RV constant \\  
              & Ca      & 116\,072     & B2V         & B2\,V             & New SB1 with preliminary orbit \\  
\\
15185--4753   & Ca      & 135\,748     & 0.13        & A2\,V             & New SB1 with preliminary orbit \\    
\\
20118--6337   & Aa+Ab+B & 191\,056     & A0V+...     & A1\,V+A1\,V        & New SB2 (Aab) with orbit\\    
              & C       &            & 0.09        & A1\,V             & RV constant \\       
\hline
\noalign{\smallskip}
\end{tabular} 

\end{table*}

\onltab{2}{
\begin{table}
\caption{Radial velocities of constant stars.}\label{tab.rv_const}
\centering
\begin{tabular}{cccrr}
\hline \hline
\noalign{\smallskip}
\multicolumn{1}{c}{WDS}&
\multicolumn{1}{c}{Subsystem}&
\multicolumn{1}{c}{HJD}    &  
\multicolumn{1}{c}{RV} & 
\multicolumn{1}{c}{Err} \\
&&&km\,s$^{-1}$&km\,s$^{-1}$\\ \hline
\noalign{\smallskip}
08314-3904      &C      &2454545.7377   &-13.4   &0.6\\
08314-3904      &C      &2454757.8605   &-13.1   &0.4\\
08314-3904      &C      &2454911.6648   &-13.7   &0.4\\
10209-5603      &C      &2454522.7974   & 6.5    &0.8\\
10209-5603      &C      &2454908.7462   & 5.8    &0.5\\
10209-5603      &C      &2454912.7841   & 5.5    &0.5\\
10209-5603      &C      &2454960.5717   & 5.7    &0.5\\
10209-5603      &C      &2454963.6431   & 7.2    &0.7\\
10209-5603      &C      &2455288.6694   & 5.6    &0.4\\
13226-6059      &AB     &2454518.7815   &-6.0    &9.1\\ 
13226-6059      &AB     &2454908.7748   &-0.4    &10.9\\
13226-6059      &AB     &2454910.8490   &-3.4    &8.2\\
13226-6059      &AB     &2454960.6387   &-5.3    &9.1\\
13226-6059      &AB     &2454962.6180   &-2.7    &8.2\\
13226-6059      &AB     &2456486.4934   &-1.3    &7.9\\
13226-6059      &AB     &2456487.4744   & 4.5    &8.1\\
20118-6337      &C      &2454572.8787   &-10.3   &5.3\\
20118-6337      &C      &2454755.5929   & -6.2   &6.1\\
20118-6337      &C      &2454961.8300   & -2.6   &5.7\\
20118-6337      &C      &2454962.8404   &-13.0   &5.7\\
20118-6337      &C      &2454963.8644   & -1.9   &5.2\\
20118-6337      &C      &2455084.6463   &-14.1   &6.2\\
20118-6337      &C      &2456486.6993   &-10.4   &5.8\\
20118-6337      &C      &2456486.8256   &-4.6    &5.1\\
20118-6337      &C      &2456487.6327   &-7.3    &5.8\\
20118-6337      &C      &2456487.7556   &-9.9    &4.9\\
20118-6337      &C      &2456488.6383   &-7.8    &5.1\\
20118-6337      &C      &2456488.7647   &-9.3    &6.0\\
\hline
\end{tabular}
\end{table}
}

\onltab{3}{
\begin{table*}
\caption{Radial velocities of variable stars.}\label{tab.rv_var}
\centering
\begin{tabular}{ccccrr}
\hline \hline
\noalign{\smallskip}
\multicolumn{1}{c}{WDS}&
\multicolumn{1}{c}{Subsystem}&
\multicolumn{1}{c}{HJD}    &  
\multicolumn{1}{c}{Phase}    &  
\multicolumn{1}{c}{RV} & 
\multicolumn{1}{c}{Err} \\
&&&&km\,s$^{-1}$&km\,s$^{-1}$\\ \hline
\noalign{\smallskip}
08314-3904	&B	&2454520.6609	&      &16.7	&1.9\\
08314-3904	&B	&2454545.6802	&      &10.3	&3.3\\
08314-3904	&B	&2454757.8324	&      &-1.3	&1.7\\
08314-3904	&B	&2454908.6405	&      &23.1	&2.4\\
08314-3904	&B	&2454911.6338	&      &19.9	&1.5\\
08314-3904	&B	&2454960.5078	&      &-2.8	&3.4\\
08314-3904	&B	&2455289.5845	&      &34.5	&3.2\\
08314-3904	&B	&2456336.7409	&      &24.7	&4.7\\
08314-3904	&B	&2456337.5494	&      &12.0	&4.3\\
08314-3904	&B	&2456338.7387	&      &15.3	&5.1\\
10209-5603      &A      &2454519.7156   &      &14.1    &3.0\\
10209-5603      &A      &2454544.7909   &      &13.0    &3.3\\
10209-5603      &A      &2454908.7114   &      &12.6    &3.2\\
10209-5603      &A      &2454911.7532   &      &10.4    &3.0\\
10209-5603      &A      &2454960.5315   &      &12.1    &3.1\\
10209-5603      &A      &2454962.5775   &      &13.4    &3.0\\
10209-5603      &A      &2455288.6449   &      &15.9    &3.2\\
10209-5603      &A      &2455291.7027   &      &11.6    &3.3\\
10209-5603      &A      &2456487.4665   &      &10.6    &3.0\\
\hline
\end{tabular}
\end{table*}
}

\onltab{4}{
\begin{table*}
\caption{Radial velocities of single-lined spectroscopic binaries.}\label{tab.rv_sb1}
\centering
\begin{tabular}{ccccrr}
\hline \hline
\noalign{\smallskip}
\multicolumn{1}{c}{WDS}&
\multicolumn{1}{c}{Subsystem}&
\multicolumn{1}{c}{HJD}    &  
\multicolumn{1}{c}{Phase}    &  
\multicolumn{1}{c}{RV} & 
\multicolumn{1}{c}{Err} \\
&&&&km\,s$^{-1}$&km\,s$^{-1}$\\ \hline
\noalign{\smallskip}
08314-3904	&Aa	&2454519.6730	&0.0253 &4.5	&1.4\\
08314-3904	&Aa	&2454545.6685	&0.3801	&4.3	&1.4\\
08314-3904	&Aa	&2454757.8231	&0.2761 &-17.7	&1.4\\
08314-3904	&Aa	&2454908.6304	&0.3346	&-2.9	&1.4\\
08314-3904	&Aa	&2454911.6235	&0.3755 &3.2	&1.4\\
08314-3904	&Aa	&2454960.4976	&0.0426 &3.4	&1.4\\
08314-3904	&Aa	&2455288.5632	&0.5208	&32.4	&1.4\\
08314-3904	&Aa	&2455291.6589	&0.5631	&35.9	&1.4\\
08314-3904	&Aa	&2456336.7289	&0.8285	&35.5	&1.4\\
08314-3904	&Aa	&2456337.5401	&0.8396 &33.2	&1.4\\
08314-3904	&Aa	&2456338.7264	&0.8558	&34.7	&1.4\\
08314-3904	&Aa	&2456696.5787   &0.7406 &43.52  &1.4\\
13226-6059      &Ca     &2454518.7860   &0.1221 &-34.2  &7.7\\ 
13226-6059      &Ca     &2454544.8437   &0.8859 & -0.3  &9.6\\  
13226-6059      &Ca     &2454908.7781   &0.3526 &-32.8  &9.0\\
13226-6059      &Ca     &2454910.8519   &0.8909 & 14.8  &7.4\\
13226-6059      &Ca     &2454960.6417   &0.8148 &  7.5  &7.2\\
13226-6059      &Ca     &2454962.6213   &0.3287 &-30.6  &8.7\\
13226-6059      &Ca     &2455288.7134   &0.9725 & -6.2  &8.4\\
13226-6059      &Ca     &2455289.7776   &0.2488 &-23.7  &7.0\\
13226-6059      &Ca     &2455291.7775   &0.7679 &  9.9  &9.9\\
13226-6059      &Ca     &2456336.8834   &0.0465 &-13.4  &7.1\\
13226-6059      &Ca     &2456337.8781   &0.3047 &-30.9  &8.9\\
13226-6059      &Ca     &2456338.8414   &0.5548 &-21.4  &8.8\\
13226-6059      &Ca     &2456486.4971   &0.8818 & -3.6  &6.8\\
13226-6059      &Ca     &2456486.5893   &0.9057 &  1.1  &7.3\\
13226-6059      &Ca     &2456487.4777   &0.1363 &-28.3  &8.6\\
13226-6059      &Ca     &2456487.5907   &0.1657 &-29.1  &8.9\\
13226-6059      &Ca     &2456488.5182   &0.4064 &-36.7  &6.8\\
13226-6059      &Ca     &2456488.5905   &0.4252 &-31.1  &7.2\\
15185-4753      &Ca     &2454519.8976   &0.7531 &  8.5  &1.0\\
15185-4753      &Ca     &2454520.8523   &0.8303 & 10.3  &0.6\\
15185-4753      &Ca     &2454544.8646   &0.7739 &  9.6  &0.7\\
15185-4753      &Ca     &2454908.8330   &0.2341 &  4.8  &0.6\\
15185-4753      &Ca     &2454911.7887   &0.4734 &  4.7  &0.8\\
15185-4753      &Ca     &2454960.7024   &0.4325 &  5.9  &0.6\\
15185-4753      &Ca     &2454962.6509   &0.5902 &  6.3  &0.7\\
15185-4753      &Ca     &2455083.5063   &0.3725 &  5.3  &0.8\\
15185-4753      &Ca     &2455084.5086   &0.4536 &  5.8  &0.7\\
15185-4753      &Ca     &2455086.4888   &0.6139 &  5.1  &0.7\\
15185-4753      &Ca     &2455087.4896   &0.6949 &  7.6  &0.6\\
15185-4753      &Ca     &2456486.5453   &0.9368 &  7.3  &0.6\\
15185-4753      &Ca     &2456486.6458   &0.9449 &  8.7  &0.7\\
15185-4753      &Ca     &2456487.5409   &0.0174 &  7.6  &0.7\\
15185-4753      &Ca     &2456487.6587   &0.0269 &  6.8  &0.8\\
15185-4753      &Ca     &2456488.5337   &0.0977 &  7.5  &0.6\\
15185-4753      &Ca     &2456488.6623   &0.1081 &  7.5  &0.7\\
\hline
\end{tabular}
\end{table*}
}

\onltab{5}{
\begin{table*}
\caption{Radial velocities of double-lined spectroscopic binaries.}\label{tab.rv_sb2}
\centering
\begin{tabular}{ccccrrrr}
\hline \hline
\noalign{\smallskip}
\multicolumn{1}{c}{WDS}&
\multicolumn{1}{c}{Subsystem}&
\multicolumn{1}{c}{HJD}    &  
\multicolumn{1}{c}{Phase}    &  
\multicolumn{1}{c}{RV$_{\rm A}$} & 
\multicolumn{1}{c}{Err$_{\rm A}$} &
\multicolumn{1}{c}{RV$_{\rm B}$} & 
\multicolumn{1}{c}{Err$_{\rm B}$} \\
\multicolumn{1}{c}{}&
\multicolumn{1}{c}{}&
\multicolumn{1}{c}{}&
\multicolumn{1}{c}{}&
\multicolumn{1}{c}{km\,s$^{-1}$}&
\multicolumn{1}{c}{km\,s$^{-1}$}&
\multicolumn{1}{c}{km\,s$^{-1}$}&
\multicolumn{1}{c}{km\,s$^{-1}$}\\ \hline
\noalign{\smallskip}
08079-6837	&Aab	&2454519.7231	&0.3456	&-49.8	&1.5	&88.6	&2.0\\
08079-6837	&Aab	&2454520.6751	&0.4128	&-28.7	&1.5	&59.6	&2.0\\
08079-6837	&Aab	&2454544.5800	&0.0999	&-32.5	&1.5	&70.8	&2.0\\
08079-6837	&Aab	&2454545.6468	&0.1752	&-53.3	&1.5	&97.8	&2.0\\
08079-6837	&Aab	&2454546.6365	&0.2450	&-61.9	&1.5	&105.9	&2.0\\
08079-6837	&Aab	&2454572.6090	&0.0780	&-27.0	&1.5	&56.9	&2.0\\
08079-6837	&Aab	&2454573.5823 	&0.1466	&-47.8	&1.5	&83.1	&2.0\\
08079-6837	&Aab	&2454574.5629	&0.2158	&-58.6	&1.5	&97.2	&2.0\\
08079-6837	&Aab	&2454576.6181	&0.3609 &-44.5	&1.5	&85.5	&2.0\\
08079-6837	&Aab	&2454577.6017	&0.4303	&-23.1	&1.5	&54.9	&2.0\\
08079-6837	&Aab	&2454579.5294	&0.5663 &32.4 	&1.5	&-34.7	&2.0\\
08079-6837	&Aab	&2454908.6100	&0.7905	&66.7	&1.5	&-95.5	&2.0\\
08079-6837	&Aab	&2454909.6216	&0.8619	&56.0	&1.5	&-72.0	&2.0\\
08079-6837	&Aab	&2454910.6880	&0.9371 &29.3	&1.5	&-33.2	&2.0\\
08079-6837	&Aab	&2454911.5975	&0.0013	&6.5	&2.5	&5.89	&3.0\\
08079-6837	&Aab	&2454912.6115	&0.0729	&-27.8	&1.5	&48.7 	&2.0\\
08079-6837	&Aab	&2454960.4905	&0.4518	&-11.7	&1.5	&38.8	&2.0\\
08079-6837	&Aab	&2454961.4652	&0.5206	&16.2	&2.5	&-5.1	&3.0\\
08079-6837	&Aab	&2454962.5361	&0.5962	&45.5	&1.5	&-48.1	&2.0\\
08079-6837	&Aab	&2454963.5350	&0.6667	&66.6	&1.5	&-77.8	&2.0\\
08079-6837	&Aab	&2455288.5548	&0.6043	&48.9	&1.5	&-53.8	&2.0\\
08079-6837	&Aab	&2455289.5474	&0.6743	&63.8	&1.5	&-80.2	&2.0\\
08079-6837	&Aab	&2455291.6380	&0.8218 &64.0	&1.5	&-82.0	&2.0\\
08079-6837	&Aab	&2455529.7698	&0.6275	&55.2	&1.5	&-67.3	&2.0\\
08079-6837	&Aab	&2456336.5230	&0.5623 &32.8	&1.5	&-30.7	&2.0\\
08079-6837	&Aab	&2456338.6493	&0.7124	&71.3	&1.5	&-88.7	&2.0\\
08079-6837	&Bab	&2454544.5861	&	&47.7	&2.4    &-45.3	&2.5\\
08079-6837	&Bab	&2454908.6159	&	&-52.4	&1.1    &62.2	&1.2\\
08079-6837	&Bab	&2454912.6190	&	&-13.3	&2.1    &19.9	&2.0\\
08079-6837	&Bab	&2454963.5418	&	&41.9	&2.3    &-40.3	&1.7\\
10209-5603      &Bab	&2454520.7560   &	&84.9   &2.1    &-99.8 	&2.2\\
10209-5603      &Bab	&2454573.6548   &	&-42.8	&2.5    &96.4 	&3.0\\
10209-5603      &Bab	&2454908.7216   &	&62.1	&2.8    &-71.7  &2.4\\
10209-5603      &Bab	&2454960.5417   &	&79.0	&1.7    &-98.1 	&2.0\\
10209-5603      &Bab	&2455292.7439   &	&-7.1	&1.5    &38.0 	&2.3\\
10209-5603      &Bab	&2456336.8178   &	&78.9	&2.9    &-110.7 &4.7\\
10209-5603      &Bab	&2456338.8573   &	&-68.3	&2.2    & 122.7	&2.7\\
20118-6337      &Aab	&2454574.8617   &0.8460	& 56.9	&1.9    &-58.1	&1.1    \\
20118-6337      &Aab	&2454754.5853   &0.5966	&39.7 	&1.1    &-39.7	&1.1    \\
20118-6337      &Aab	&2454961.7888   &0.3179	&-61.0	&1.7    &58.5	&1.6    \\
20118-6337      &Aab	&2454963.8326   &0.1338	&-51.9 	&1.5    &49.7	&0.9    \\
20118-6337      &Aab	&2455083.6507   &0.9685	& 8.0	&2.0    &-20.3	&2.0    \\
20118-6337      &Aab	&2455084.6143   &0.3532	&-58.9	&1.7    &52.7	&1.5    \\
20118-6337      &Aab	&2455085.6577   &0.7697	&65.9 	&1.5    &-69.3	&1.4    \\
20118-6337      &Aab	&2455086.6321   &0.1587	&-60.1	&1.7    &54.3	&2.2    \\
20118-6337      &Aab	&2455087.6014   &0.5457	& 16.6	&1.1    &-24.3	&0.8    \\
20118-6337      &Aab	&2455288.8592   &0.8933	&41.5 	&1.4    &-43.4	&1.0    \\
20118-6337      &Aab	&2455291.8726   &0.0963	&-40.0 	&1.0    &38.1	&1.2    \\
20118-6337      &Aab	&2456486.6665   &0.0906	&-40.1 	&1.2    &38.8	&1.6    \\
20118-6337      &Aab	&2456486.8041   &0.1456	&-54.9 	&1.3    &56.4	&1.4    \\
20118-6337      &Aab	&2456487.6096   &0.4671	&-18.7	&1.6    &9.4	&1.1    \\
20118-6337      &Aab	&2456487.7811   &0.5356	& 11.9	&1.7    &-18.9	&1.8    \\
20118-6337      &Aab	&2456488.6104   &0.8667	& 50.1	&1.5    &-53.8	&1.0    \\
20118-6337      &Aab	&2456488.7421   &0.9193	& 32.5	&1.7    &-36.3	&1.1    \\
\hline
\end{tabular}
\end{table*}
}
From the analysis of our spectroscopic observations and measured RVs, we concluded that some of the studied systems are formed by more components than the previously known. In some cases the new components were visible in the spectra, while in others their existence was detected through RV variations. 
Particularly in this paper we present the study of six multiple systems classified as triples in MSC, in which we detected new components. Each of these will be discussed in detail below. Table~\ref{table:1} summarizes the results obtained in each case. The observed components are listed in column 2 and identified by their HD number in column 3. For each component, column 4 lists the published spectral type or, if not available, the $B-V$ color index, and column 5 presents the spectral type determined in this work. Finally, a brief description of the results is given in column 6.

The measured RVs of the 13 observed components in the six systems analysed in this work are listed in Tables 2 to 5, which correspond to RV constant stars, RV variables, single-lined, and double-lined spectroscopic binaries, respectively.

\subsection{WDS~08079--6837}

In this system, the components A and B have a separation of 6.1$\arcsec$ and a magnitude difference $\Delta V=2.93$. Star A (HD~68\,520) is catalogued as a single-lined spectroscopic binary, for which \citet{1915LicOB...8..127S} computed an orbit with a period of 14.1683 days. \citet{2000IBVS.4827....1M} detected the secondary component of this subsystem in some observations and determined a mass ratio $M_\mathrm{1}/M_\mathrm{2}=1.30$. Recently, \citet{2012MNRAS.424.1925C} obtained two spectra of this binary, confirming its status as SB2. 

The subsystem A was observed frequently, with the aim of detecting spectral features of the secondary and computing orbital parameters for both components. In RV measurements we employed synthetic templates with $T_{\rm eff}=15000\:\rm{K}$ and $T_{\rm eff}=13000\:\rm{K}$.
We obtained 26 spectra of this binary. We reconstructed the spectra of the components combining 24 observed spectra, since in the other two the spectral lines of both components were too blended to measure preliminary RVs. However, in a subsequent calculation with the method GL06, we measured RVs for both components even in the two heavily blended spectra. 
The reconstructed spectra of the components led to confirm the spectral type B6\,IV published in the MSC for the primary and to estimate a spectral type B8 for the secondary.  

The period by \citet{1915LicOB...8..127S} represents well our RV data so we take it as the starting  value of period for the orbital elements computation. With all 7 parameters free, the orbital fitting gave an eccentricity indistinguishable from zero. Therefore, we adopted a circular orbit and we determined final values for the others orbital elements (except $\omega$). These parameters are listed in Table~\ref{table:6}, which includes also the number of spectra (N$_{\mathrm{sp}}$) and rms deviation of the velocities ($\sigma_\mathrm{RVa}$ and $\sigma_\mathrm{RVb}$).

\begin{figure}
   \centering
   \includegraphics[width=0.95\linewidth]{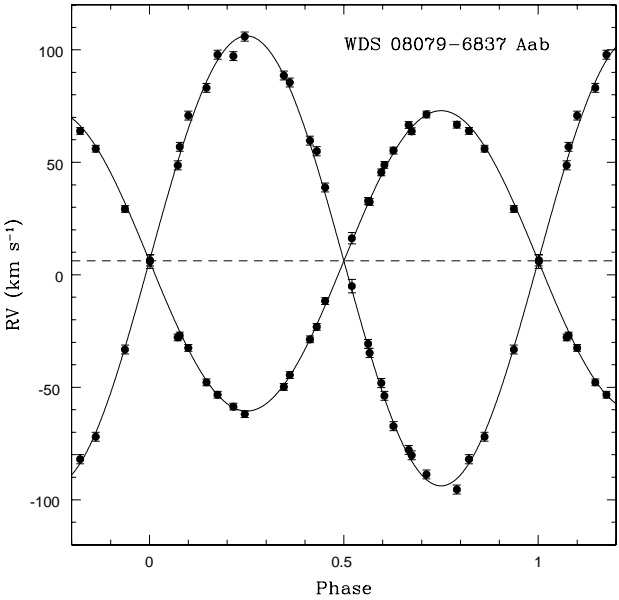}
      \caption{Radial velocity curves of WDS~08079--6837~Aab. Filled circles represent our measurements. Continuous line shows the orbital fitting. Dashed line indicates the barycentre's velocity.}
         \label{fig1}
   \end{figure}

We obtained a mass ratio $q=0.67$, lower than the one given by Medici \& Hubrig ($q=0.77$) and higher than the value estimated by Sanford ($q=0.23$). However, there is a good agreement between the parameters found in this study and those calculated by Sanford for the orbit of the primary component Aa. Figure~\ref{fig1} presents the fitting of the RVs of both components.

\begin{figure}
   \centering
   \includegraphics[width=0.95\linewidth]{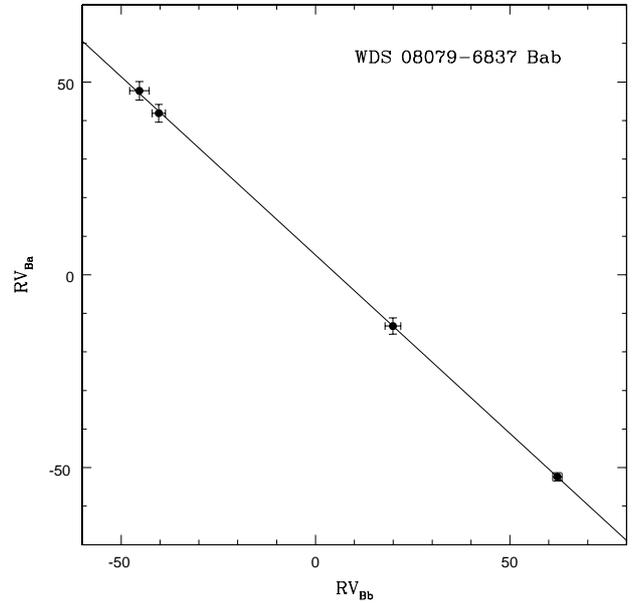}
      \caption{Linear correlation between the RVs of the components of WDS~08079--6837~Bab. Continuous line shows the fitting corresponding to $\mathrm{RV}_\mathrm{Ba}=-0.93\;\mathrm{RV}_\mathrm{Bb}+5.1\;{\rm km\:s}^{-1}$, 
which leads to $q=0.93\pm0.03$ and $V_{\gamma}=2.7\pm0.7\;{\rm km\:s}^{-1}$. 
Filled circles represent our measurements.}
         \label{fig2}
   \end{figure}

It was difficult to observe the component B of this multiple system due to the proximity of star A, which is about 15 times brighter. Thus, we only obtained six useful spectra of B, which showed morphological variations indicating two spectral components. 
We applied cross-correlations with an observed template of spectral type A1V. In only four spectra we measured RVs of both components separately. In the other two we obtained only one broadened peak in the cross-correlation function. 
There is a linear correlation between the RVs measured for both components, which confirms that they form a double-lined spectroscopic binary. The linear fitting showed in Figure~\ref{fig2} leads to a mass ratio $q=0.93\pm0.03$ and a barycentre's velocity $V_{\gamma}=2.7\pm0.7\;{\rm km\:s}^{-1}$. 
The small number of observations does not allow to derive the orbital period, but given the RV variations detected within
the same observing run, it would be of the order of a few days.

In conclusion, we found that the component B is a binary subsystem and we obtained the orbital parameters of Aab, through the fitting of the RVs of its two components. Furthermore, there is a good agreement between the velocities obtained for the barycentres of both subsystems, which confirms that they are physically bound to each other.
Therefore, the system WDS~08079--6837, previously classified as a triple system, is in fact a quadruple.

\subsection{WDS~08314--3904}

This system is formed by components A and B separated by 4.47$\arcsec$, and a third component C at 30.8$\arcsec$ from AB. In the MSC, the physical link between the components has been established on the basis of the criterion of hypothetical dynamical parallax and the similarity between the distances estimated from the spectral type and apparent magnitude.

This multiple was observed in eight observing runs, in which we obtained 12 spectra of A (HD~72\,436). We did not detect any spectral feature indicating the presence of other spectroscopic component. Radial velocities,  measured using a template of spectral type B4V convolved with a rotational profile  of $v\sin i=150\;{\rm km\:s}^{-1}$, showed to be variable.

\begin{figure}
   \centering
   \includegraphics[width=0.90\linewidth]{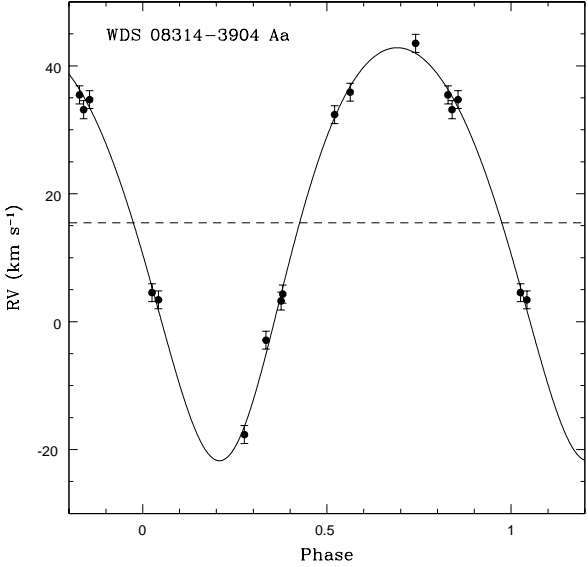}
      \caption{Radial velocity curve of WDS~08314--3904~Aa. Filled circles represent our measurements. Continuous line shows the orbital fitting. Dashed line indicates the barycentre's velocity.}
         \label{fig3}
   \end{figure}

The best fitting of the RV data was achieved for the eccentric orbit showed in Figure~\ref{fig3} and Table~\ref{table:7}. The fitting of the data is good: the rms deviations are similar to the measurement errors. However, we classified this orbit as ``preliminary'' since the number of measurements is small and additional observations are required to confirm these orbital parameters.

We obtained 10 spectra of the component B, in which we measured RVs through cross-correlations with an observed template of spectral type B6V and $v\sin i=150\;{\rm km\:s}^{-1}$. The measurements are listed in Table~\ref{tab.rv_var}. We found a RV variation between different observing runs, although we have not yet identified the corresponding period. 

Component C is an object with low rotational velocity for which we obtained three spectra with a time baseline of approximately one year. 
The obtained RVs did not show variations, being the mean value $-13.4\pm0.2\;{\rm km\:s}^{-1}$.

The similarity between the barycentric velocity of A ($15.1\pm0.5\;{\rm km\:s}^{-1}$), and the average of the velocities measured for star B ($15.2\;{\rm km\:s}^{-1}$) supports the existence of a physical link between these two binaries. However, the velocity obtained for C differs substantially from these values, 
suggesting that this object does not belong to WDS~08314--3904. 
Consequently, the hierarchical structure of this system would be considerably different from previously
known: instead of 3 stars weakly bound to each other, it consist of two close binaries in a wide orbit. 

\begin{table*}
\caption{Definitive orbital parameters of double-lined spectroscopic subsystems.}
\label{table:6}
\centering
\begin{tabular}{c c c c c c c c c c} 
\hline 
\hline
Subsystem & N$_{\mathrm{sp}}$ & $P$ & $T_0$ & $e$ & $K_\mathrm{a}$ & $K_\mathrm{b}$ & $V_\gamma$ & $\sigma_\mathrm{RVa}$ & $\sigma_\mathrm{RVb}$ \\ 
(WDS) & & (days) & (HJD-2400000) &  & (${\rm km\:s}^{-1}$) & (${\rm km\:s}^{-1}$) & (${\rm km\:s}^{-1}$) & (${\rm km\:s}^{-1}$) & (${\rm km\: s}^{-1}$) \\
\hline 
\noalign{\smallskip}
08079--6837 A & 26 & \phantom{.}14.1697 & 54\,514.83  & 0   & \phantom{.}66.8 & 100.0 & \phantom{2}\phantom{.}6.2 & 1.8 & 3.0 \\     
             & & $\pm$0.0006 & \phantom{2}\phantom{2}\phantom{.}$\pm$0.03  &     & $\pm$0.9 & \phantom{.}$\pm$1.2 & $\pm$0.5 &     &     \\ 
\\
20118--6337 A & 17 & \phantom{2}\phantom{.}2.504836 & 54\,517.636  & 0   & \phantom{.}68.8 & \phantom{.}69.1 & \phantom{2}-2.0 & 1.9 & 2.2 \\     
             & & $\pm$0.000015 &\phantom{2}\phantom{2}\phantom{.}$\pm$0.009  &   & $\pm$1.3 & $\pm$1.3 & $\pm$0.6 &     &     \\             
\hline
\end{tabular} 
\end{table*}

\begin{table*}
\caption{Preliminary orbital parameters of single-lined spectroscopic subsystems.}
\label{table:7}
\centering
\begin{tabular}{c c c c c c c c c} 
\hline 
\hline
Subsystem & N$_{\mathrm{sp}}$ & $P$ & $T_0$ & $e$ & $\omega$ & $K_\mathrm{a}$ & $V_\gamma$ & $\sigma_\mathrm{RVa}$ \\ 
(WDS) & & (days) & (HJD-2400000) &  & (\degr) & (${\rm km\:s}^{-1}$) & (${\rm km\:s}^{-1}$) & (${\rm km\: s}^{-1}$) \\
\hline 
\noalign{\smallskip}
08314--3904 A & 12 & \phantom{.}73.25 & 55\,543.4  & \phantom{2}\phantom{.}0.155   &  190.2  & \phantom{.}32.3 & \phantom{.}15.5 & 1.4 \\    
             & & $\pm$0.04 & \phantom{2}\phantom{2}\phantom{.}$\pm$0.4  & $\pm$0.016   & \phantom{.}$\pm$9.2  & $\pm$1.0 & $\pm$0.5 &  \\  
\\
13226--6059 C & 18 & \phantom{2}\phantom{.}3.8523 &55500.71  &\phantom{2}\phantom{.}0.24  &\phantom{.}346     &\phantom{.}21.2     & -16.5    & 5.0 \\
              &    &$\pm$0.0010 &\phantom{2}\phantom{2}\phantom{.}$\pm$0.11 &$\pm$0.11 &$\pm$52 &$\pm$3.2 & $\pm$2.4 &  \\ 
\\  
15185--4753 C & 17 & \phantom{.}12.353 &55177.7 &\phantom{2}\phantom{.}0.4 &\phantom{.}324 &\phantom{2}\phantom{.}2.4 & \phantom{2}\phantom{.}6.9 &0.6 \\
              &    &$\pm$0.004 &\phantom{2}\phantom{2}\phantom{.}$\pm$0.6 &$\pm$0.1 &$\pm$18 &$\pm$0.4 &  $\pm$0.3 &  \\ 
\hline
\end{tabular} 
\end{table*}

\subsection{WDS~10209--5603}

This system has components A, B, and C catalogued in the MSC. The separation between A and B is 7.19$\arcsec$ and their published magnitude difference is $\Delta V=3.9$. Component C is located at 36.7$\arcsec$ from AB. In the catalogue, the physical link between the components has been established on the basis of the criterion of hypothetical dynamical parallax. 

\citet{1984ApJS...55..657C} classified both companions of the close visual pair and reported star B as a silicon star, giving spectral
types B4III and B9.5IV Si for stars A and B, respectively. \citet{1898ApJ.....8..116P} detected an emission in H{$\beta$} in a spectrum of component A, but this result has not been confirmed on further observations. \citet{1984A&A...134..105B} detected variations in the profiles of some spectral lines with a period of about 2.25 days. Since these variations are very similar to those observed in Be stars, HD~89\,890 is considered to belong to this class of objects but presently going through a long B-star phase \citep{2002ASPC..259..248S}. From RV measurements in 15 spectra, \citet{2003A&A...411..229R} found a period of 2.318 days, which is consistent with the data of Baade. Furthermore, \citet{1996A&A...311..579S} measured photometric variations, founding several possible periods around 4.6 days, the most prominent being 4.656 days. According with their data, they pointed out that a period of approximately 2 days is beyond their detections possibilities. 

\begin{figure}
   \centering
   \includegraphics[width=0.95\linewidth,height=8cm]{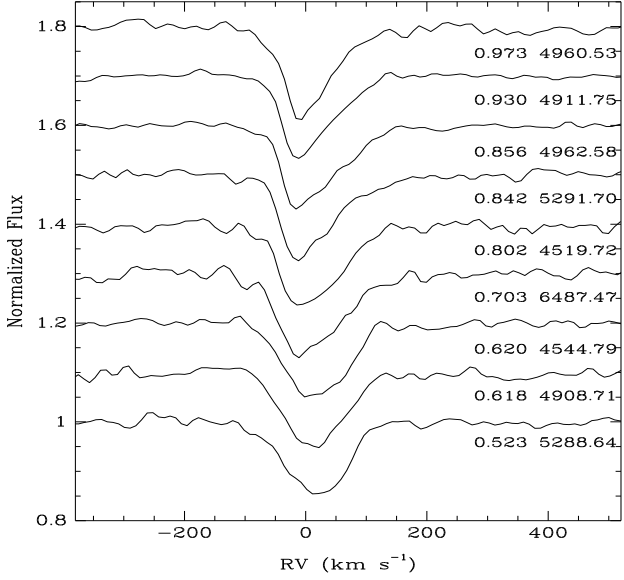}
      \caption{Profile variation of the spectral line \ion{Mg}{ii} $\lambda$4481 in HD\,89\,890.
Labels on the right  are phases (calculated with $P=2.318$) and HJD-2\,450\,000.}
        \label{plot4481}
   \end{figure}

We obtained 9 spectra of component A, which exhibit clear line-profile variations.
Figure\,\ref{plot4481} shows the line profile of the  \ion{Mg}{ii} line at $\lambda$4481. The spectra have been ordered by phases calculated using a period of 2.318 d and
a phase origin set arbitrarily at HJD\,2\,450\,000. The observed variations
are clearly consistent with the period published by \citet{2003A&A...411..229R}.
Due to the asymmetry of line profiles in HD\,89890 
we measured RVs by determining the barycentre of
the profile of individual lines. The velocities consigned
in Table~\ref{tab.rv_var} are the average of 9 metallic and \ion{He}{i} lines. 
We include this object in Table~\ref{tab.rv_var} considering that a RV variation exists, although in this case, it is not an evidence of orbital motion.

Due to the proximity of the component A, which is about 4 magnitudes brighter than B, we could only obtain seven spectra of this fainter component on the best observing nights. We detected morphological variations in the observed spectra that suggested the presence of two RV variable stars. In the spectral separation process we applied observed templates of spectral types A1V and A9IV for the RVs measurement. 
From the separated spectra we confirmed for the primary of this SB2 the classification as Silicon star given by \citet{1984ApJS...55..657C}.
The linear correlation between the velocities of both components indicated that they form an SB2 subsystem. 

\begin{figure}
   \centering
   \includegraphics[width=0.95\linewidth]{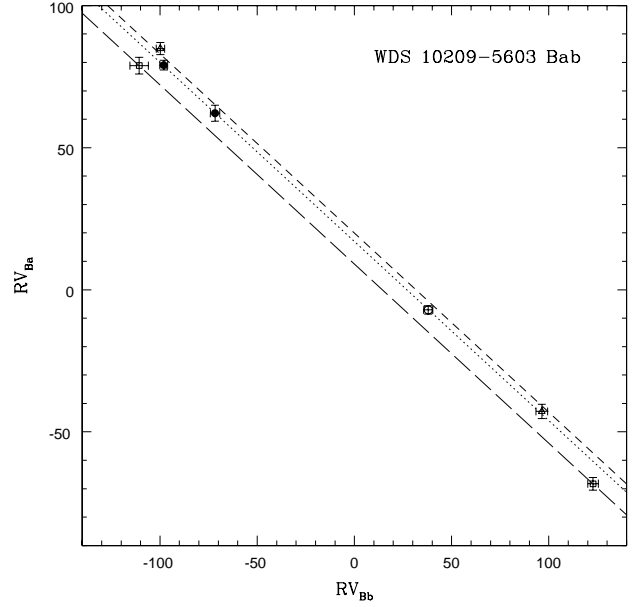}
     \caption{Linear correlation between the RVs of the components of WDS~10209--5603~Bab. Open squares represent measurements of February 2013. These are fitted by the function $\mathrm{RV}_\mathrm{Ba}=-0.63\;\mathrm{RV}_\mathrm{Bb}+9.1\;{\rm km\:s}^{-1}$, which is showed with long-dashed line. 
Open triangles represent measurements of February and April 2008, and filled circles indicate 
measurements of March and May 2009. Adopting a mass ratio $q=0.63$, these data are fitted by the function $\mathrm{RV}_\mathrm{Ba}=-0.63\;\mathrm{RV}_\mathrm{Bb}+20.0\;{\rm km\:s}^{-1}$ (short-dashed line), and the function $\mathrm{RV}_\mathrm{Ba}=-0.63\;\mathrm{RV}_\mathrm{Bb}+17.0\;{\rm km\:s}^{-1}$ (dotted line), respectively. Open circle indicates RVs measured in a spectrum of April 2010.}
 \label{fig4}
   \end{figure}

RVs of stars Ba and Bb follow the linear relation expected for the components of a binary system, however, a closer inspection of the deviations suggests a long term variation of the center-of-mass velocity. We calculated three linear fittings using only RVs measured in a same observing run or in observing runs temporally close to each other. In this way we obtained straight lines with roughly equal slopes but different intercepts, which suggests a RV variation of the barycentre of the spectroscopic binary subsystem. 
Since the available data are insufficient to determine the period of that variation, we adopted the mass ratio obtained from the fitting of the only two pairs of velocities measured in a same observing run (February 2013). From these data we obtained $q=0.63\pm0.02$ and $V_{\gamma}=5.6\pm1.5\;{\rm km\:s}^{-1}$. Adopting this value for the mass ratio, we fitted the barycentre's velocity for the others observing runs temporally close to each other. 
In Figure~\ref{fig4} we present the seven pairs of RVs measured and the fittings calculated using six of them corresponding to three different observing runs.
The center-of-mass velocity seems to have been  decreasing at a rate
of -3.5$\times 10^{-3}$ km\,s$^{-1}$\,d$^{-1}$ during the 5 years covered by our data.
The outer period would then be of some thousand days, although probably shorter than $10^4$ days since with longer periods the RV variations would be smaller than observed.
On the other hand, the relatively rapid RV variations suggest an
inner orbital period on the order of 3--4 days.
Consequently, the scarce available data suggest the existence of an intermediate hierarchical orbit between the visual pair AB and the subsystem Bab. This interesting result encourages us to continue with the observations of this subsystem, despite the difficulties due to the proximity of HD~89\,890.

We obtained six spectra of the component C with a time baseline of two years. We did not detect morphological or RV variations between them. From the comparison of these spectra with the available templates we determined a spectral type K0\,III for this object, which had not been classified previously. The RVs measured by cross correlations using a synthetic template with $T_{\rm eff}=5500\:\rm{K}$ led to a mean value $6.1\pm0.6\:{\rm km\:s}^{-1}$.

 The similarity between the mean velocity of component A, the barycentric velocities obtained for Bab, and the mean velocity found for component C supports the idea that they are gravitationally bound. With the detection of two new hierarchical levels the system
 WDS~10209--5603 would be, at least, quintuple.

According with theoretical models, a star with spectral type B3III should be significantly younger than a star of spectral type K0III if both of them have followed a normal evolution (early-B type stars evolve into supergiants KIb) . Therefore, it is striking that A and C belong to the same physical multiple system. Considering the flux ratio between these objects, estimated from our spectroscopic data and the absolute magnitudes from the \citet{Schmidt-Kaler} calibration, we found that the distance moduli of both objects are compatible with each other. This is a further evidence, in addition to the similarity of their RVs, that supports the physical link between them. 
If we assume that both objects have a common origin, then one of them should have had an abnormal evolution. 
Particularly, component A might have stayed on the main sequence over a longer period of time than a normal star with the same mass, being then a blue-straggler star. Moreover, it can be speculated that this blue-straggler could have had origin in the merger of the components of a close binary as a result of the influence of a third companion through Kozai cycles and tidal friction (see Section 4) \citep{2009ApJ...697.1048P}. The profile variations of its spectral lines limits our detection possibilities of a possible orbital motion only up to periods of a few tens of days. Therefore, we do not rule out the existence of an additional companion of A in a hierarchical level below the visual pair AB.

\subsection{WDS~13226--6059}

The components A and B of this multiple system form a visual subsystem with a separation of 0.17$\arcsec$. In the first hierarchical level, the pair AB (HD~116\,087) and the component C (HD~116\,072) form a common proper motion and RV pair, separated by 60$\arcsec$. We observed AB and C during four and seven observing runs, respectively, with a time baseline of 5.4 years.

We took seven spectra of the pair AB, in which a single set of broad spectral lines is visible, without 
significant morphological variations. In cross-correlations we employed a synthetic template with $T_{\rm eff}=18000\:\rm{K}$ and $v\sin i=200\;{\rm km\:s}^{-1}$. We did not measure RV variations greater than the errors given by the task {\itshape fxcor}, which were around a mean value of 8.8 ${\rm km\:s}^{-1}$. We obtained a mean RV $-2.1\pm3.5\;{\rm km\:s}^{-1}$.

We obtained 18 spectra of the component C. In the measurement of RVs we employed a synthetic template with $T_{\rm eff}=19000\:\rm{K}$ and $v\sin i=200\;{\rm km\:s}^{-1}$. Although we obtained good correlation peaks, the measurement errors were about 8.0 ${\rm km\:s}^{-1}$, due to its high rotation. The measurements revealed a RV variation, but we did not detect spectral features of a companion. By using $pdm$ we obtained tentative periods of 3.85 and 4.537 days. The least squares fitting of the RV curve led to a slightly lower rms deviation for the former. However, the quality of the orbit is poor because of the low amplitude-to-error ratio of the RV curve, so we do not discard the another possible period. The  parameters of the preliminary orbit are listed in Table~\ref{table:6}. Parameter uncertainties are significant since RV errors are only $\sim$3 times smaller than the semiamplitude.

The orbital period is relatively short while the calculated eccentricity could be considered significantly different from zero: $e/\sigma_e\sim 2$.
A statistical test of the null eccentricity hypothesis \citep[as formulated by][]{1971AJ...76..544L} gives a probability of only 5\% that the orbit is circular. However, we noted that the rms of residuals (~5 km\,s$^{-1}$) is somewhat smaller than the formal measurement errors (~8 km\,s$^{-1}$).
If parameter errors are calculated using this value instead of the rms of the deviations from the fits, the probability of a circular orbit is about 30\%.
Even if the discrepancy between rms deviations and formal errors is due probably to the overestimation of the latter, we consider that this orbit and particularly its eccentric character should be taken with caution until a large number of observations allow to obtain a more reliable orbit.

Although the mean velocity of AB and the barycentric velocity obtained for Ca do not agree within the errors, for the moment we do not rule out the existence of a physical link between them, at least until  the orbital parameters of Ca are confirmed. Therefore WDS~13226--6059 is, at least, a quadruple system.

\begin{figure}
   \centering
   \includegraphics[width=0.95\linewidth]{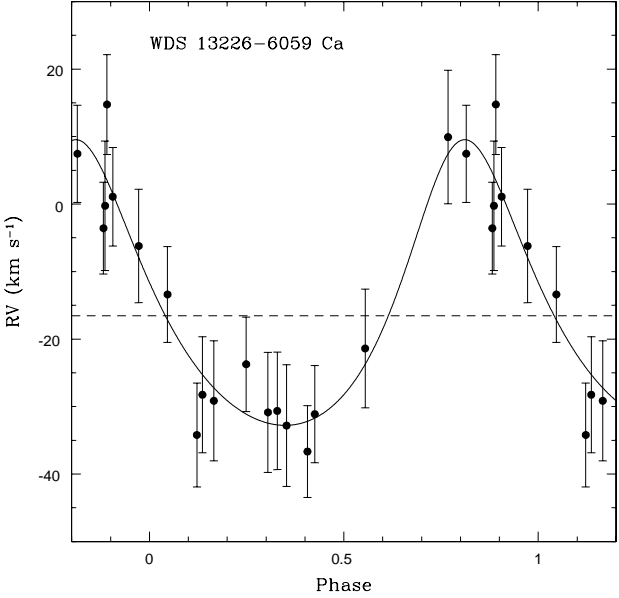}
      \caption{Radial velocity curves of WDS~13226--6059~Ca. Filled circles represent our measurements. Continuous line shows the orbital fitting. Dashed line indicates the barycentre's velocity.}
         \label{fig5}
   \end{figure}

\subsection{WDS~15185--4753}

This system has components A, B and C catalogued in the MSC. A and B form a visual pair with a separation of 1.1$\arcsec$, without published orbit. The component C is at 22.6$\arcsec$ from AB and forms a common proper motion system with this pair. Based on infrared observations, \citet{2001A&A...372..152H} reported the detection of an additional component located 6.15$\arcsec$ apart from A. From the $K$ magnitude they estimate that this fourth component is a pre-main sequence star with a mass of approximately 0.05 M$_{\sun}$. Given the very low mass inferred, Hubrig et al. pointed out that this faint object could be a brown dwarf.  

The subsystem AB has many previous spectroscopic studies and its published spectral type is B8Ve. Based on observations of high angular resolution ($\sim0.1\arcsec$) with adaptive optics, \citet{2010MNRAS.405.2439O} reported the detection of a visual companion of the component A, located at 0.98$\arcsec$ and position angle of 305$\degr$. However, this is probably the visual B companion catalogued, which, according with the MSC, is at a position angle of 311$\degr$.  

We carried out observations of the component C (HD~135\,748), which had not detailed spectroscopic studies. We obtained 17 spectra during six observing runs with a time baseline of 5.4 years.  From our spectra, we determine a spectral type A2\,V. To measure RVs we employed a template with $T_{\rm eff}=8500\:\rm{K}$ and $v\sin i=30\;{\rm km\:s}^{-1}$. The measurement errors were around 0.7 ${\rm km\:s}^{-1}$ and detected RV variations of a few kilometers per second (Table~\ref{tab.rv_sb1}). 

\begin{figure}
   \centering
   \includegraphics[width=0.95\linewidth]{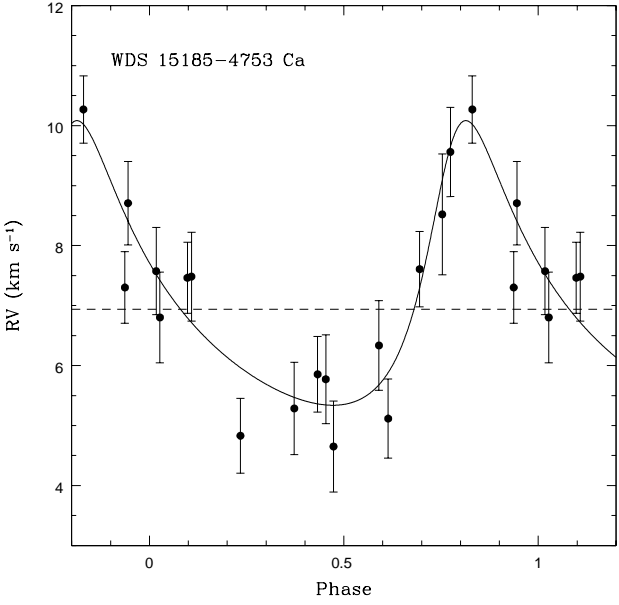}
      \caption{Radial velocity curves of WDS~15185--4753~Ca. Filled circles represent our measurements. Continuous line shows the orbital fitting. Dashed line indicates the barycentre's velocity.}
         \label{fig6}
   \end{figure}

The best fitting of our RVs (Figure~\ref{fig6}) was obtained for an eccentric orbit whose parameters are listed in Table~\ref{table:7}. However, it should be mentioned that another possible orbit, with a period of about 2.69 days, is also consistent with the observational data. 
We consider therefore, these orbital parameters to be very preliminary. However,
the classification of C as a single-lined spectroscopic binary is a robust result: 
the reduced chi squared is $\chi^2/(n-1)$ = 4.85.

We took a spectrum of HD~135\,734 (AB subsystem). We detected two overlapping line sets corresponding to objects of similar spectral type and RV but very different projected rotational velocity. \citet{2006yCat..73710252L} measured for HD~135\,734 a projected RV  of $280\;\pm20\;{\rm km\:s}^{-1}$, in good agreement with the value previously measured by \citet{2001yCat..33780861C} ($v\sin i=278\;\pm25\;{\rm km\:s}^{-1}$). 
Those results are consistent with the broad line component detected in our spectra, while the other has $v\sin i\approx50\;{\rm km\:s}^{-1}$. In cross-correlations we employed a template with spectral type B7V convolved with a rotational profile corresponding to the object with lower rotational velocity. In this way, we measured a RV $6.8\pm1.8\;{\rm km\:s}^{-1}$. This value agrees with the barycentric velocity of Ca, which confirms the physical link between both subsystems. Therefore, WDS~15185--4753 has, at least, four components, although if the gravitational link between the infrared source detected by \citet{2001A&A...372..152H} and the pair AB is confirmed, this would be a quintuple system.

\subsection{WDS~20118--6337}

This multiple is formed by a visual pair AB (HD~191\,056) with a separation of 0.3$\arcsec$ and a third component C located at 7.34$\arcsec$ from AB. In the MSC, the physical link between the components has been established on the basis of the criterion of hypothetical dynamical parallax and the proximity between the distances estimated from the spectral type and apparent magnitude.

We obtained 17 spectra of the subsystem AB, with a time baseline of 5.2 years. At first sight, we detected two components of different brightness showing RV variations of short period. 
In spectra in which the lines of these two components are well separated we detected the spectral features of a third object, although in all cases these features were blended with the lines of the two brightest components. 

\begin{figure}
   \centering
   \includegraphics[width=0.95\linewidth]{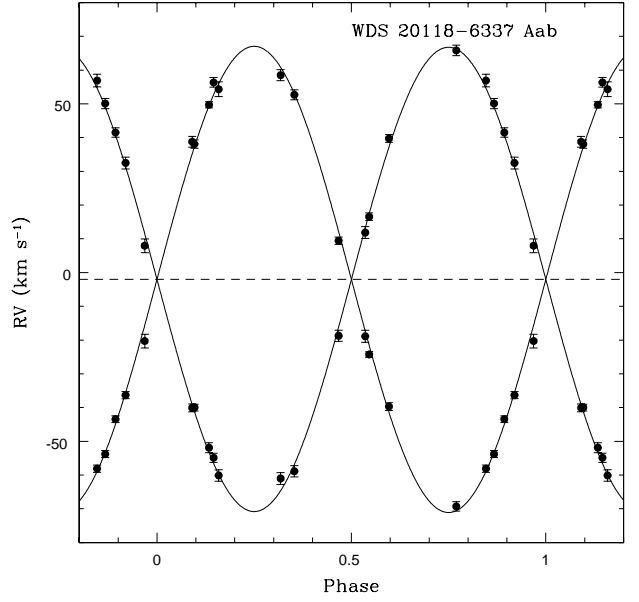}
      \caption{Radial velocity curves of WDS~20118--6337~Aab. Filled circles represent our measurements. Continuous line shows the orbital fitting. Dashed line indicates the barycentre's velocity.}
         \label{fig7}
   \end{figure}

We measured RV of the two brightest components through cross-correlations, using a template of spectral type A1V convolved with a rotational profile corresponding to $v\sin i=25\;{\rm km\:s}^{-1}$. We found that the velocities measured correspond to two stars of equal mass forming a double-lined spectroscopic binary subsystem. 
Therefore, the difference of intensities between the spectral lines of the components is striking, especially because the component that appears brighter is not the same in all spectra. This is possibly due to the presence of a fourth star with variable RV, whose spectral features are overlapping alternately with the lines of one or another component. 
Initially, this had led to a misleading identification of the components and consequent difficulties in the period estimation of the binary subsystem. Once correctly identified both components of the SB2, we measured their RVs using the disentangling method by GL06. This technique gave good results, despite the presence of the two additional fainter components in the observed spectra. From the separated spectra, we found a spectral type A1V for both components. Figure~\ref{fig7} shows the least-squares fitting of the RVs and Table~\ref{table:6} lists the orbital parameters obtained for the SB2 subsystem. In the same way as for WDS~08079--6837~Aab, we obtained an eccentricity indistinguishable from zero, so we adopted a circular orbit and refitted the remaining orbital elements.

To sum up, based on the analysis of the spectra of the visual pair AB we detected that A is not a single component but a double-lined spectroscopic binary subsystem (Aab). We did not detect variations in its barycentre's velocity over more than five years, so the existence of a intermediate hierarchical level between Aab and the visual pair AB is not expected. We estimate that the third component detected between the lines of the SB2 is the visual B, but for the moment we cannot confirm if this has a spectroscopic companion.

We obtained 12 spectra of the component C, which does not have previous RV measurements published in the literature. We determined a spectral type A1V and  measured RVs by cross-correlations using a template with $T_{\rm eff}=9750\:\rm{K}$ and $v\sin i=130\;{\rm km\:s}^{-1}$. The measurements cover a time baseline of 5.2 years and do not show any significant RV variation. The mean RV obtained for this component is $-8.1\pm3.8\;{\rm km\:s}^{-1}$.

\section{Discussion}

\begin{figure*}
   \centering
\hfill\begin{minipage}[t]{.47\textwidth}
   \includegraphics[width=\linewidth, height=10cm, bb= 2 70 557 737]{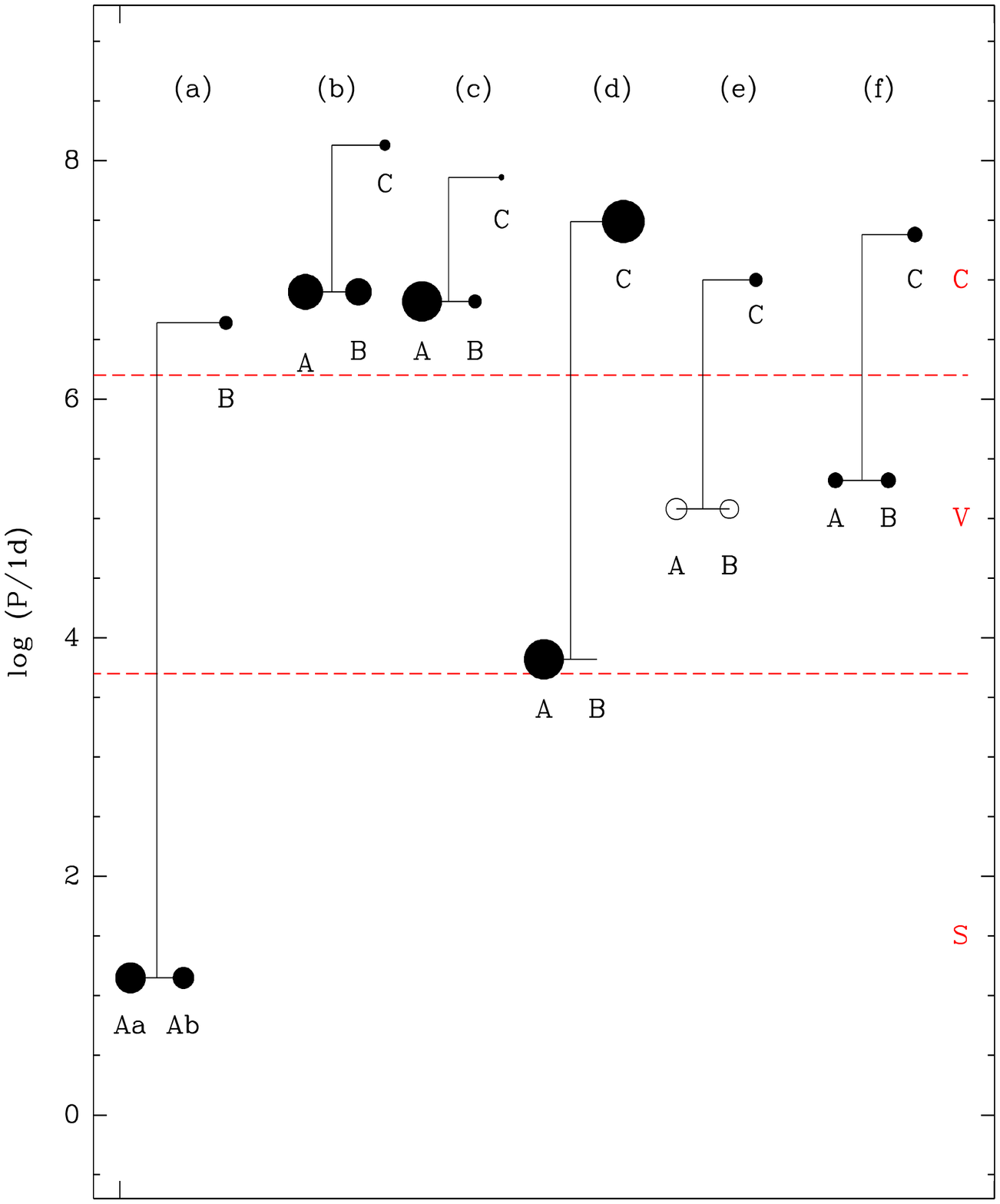}
\end{minipage}\hfill~%
\hfill\begin{minipage}[t]{.47\textwidth}
   \includegraphics[width=\linewidth, height=10cm, bb= 2 70 557 737]{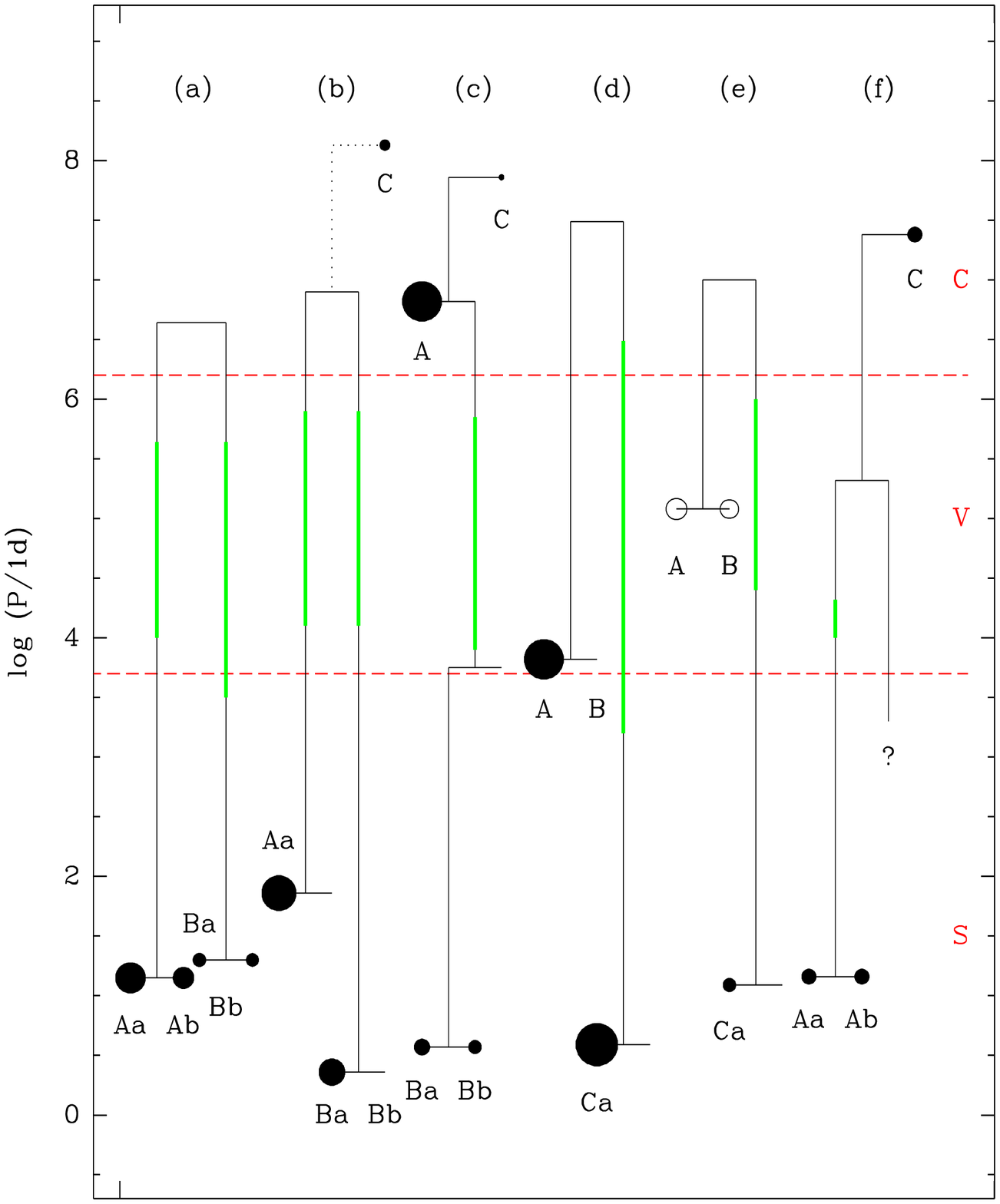}
\end{minipage}\hfill~%
      \caption{Configuration of the analysed multiple systems. {\em Left:} considering only the data catalogued in MSC. {\em Right:} including the results of this study. 
      a) WDS~08079--6837, b) WDS~08314--3904, c) WDS~10209--5603, d) WDS~13226--6059, e) WDS~15185--4753, and f) WDS~20118--6337.
Stellar components are represented by circles. Filled circles are stars observed in the present paper. No symbol is plotted for stellar components that are not visible in the spectrum (secondaries of SB1 systems). 
Dashed lines divide approximately the period axis in spectroscopic (S), visual (V), and common proper motion (C) ranges. 
Dotted lines indicate a doubtful physical link between the components. 
The size of the circles roughly scale with the stellar mass. Green thick lines mark separation intervals that have not been investigated and might harbour additional hierarchical levels.}\label{tree}
   \end{figure*}

Figure~\ref{tree} presents the hierarchical configuration of the analysed systems, considering only the data catalogued in MSC (left) and adding the results of this investigation (right). As can be seen by comparing both panels, this analysis has led to significant changes in the multiplicity order and hierarchical structure of the studied systems. 
However, it is possible that there are other components inaccessible to our observations and to the previous studies. 
Several of the studied systems present a high period ratio for orbits of consecutive hierarchical levels, which suggests that there might be undiscovered intermediate orbits.

We mark in Figure~\ref{tree} separation ranges that remain still unexplored and might harbour intermediate hierarchical levels. The maximum period that could have an intermediate orbit has been determined considering that the period ratio of adjacent orbits should be  $P_\mathrm{out}/P_\mathrm{in} \ga 10$ to maintain the stability of the system. We employed the published period of the outer orbit. 
On the other hand, for each subsystem (including spectroscopic singles) we estimated the minimum period that could have an outer undetected orbit by evaluating our variability detection limits. 
Specifically, for a given subsystem, we calculated the expected amplitude of the variations of its barycentric RV caused by an hypothetical body (assumed to have at least  0.5 M$_\odot$) moving in a circular outer orbit.
Assuming random orbital inclinations and random phases, we calculated the probability that
during the time span of our observations the RV variations of the observed inner subsystem are larger than the corresponding measurement errors.
These calculations were performed as a function of the outer period, so that we finally obtained the
longest period for which we have at least a 50\% detection probability. 

Even though this spectroscopic study has  contributed significantly to the knowledge of the hierarchical structure of the studied systems, in all cases it would be interesting to extend the search of additional components to the range of visual separations through high-angular resolution observations. This could contribute not only to the detection of intermediate orbits but also to the detection of long-period companions of components classified as single stars. Particularly, the components A and C of WDS~10209--5603 and the component C of WDS~20118--6337 could have additional components in intermediate orbits with periods greater than approximately $10^4$ days that would have gone unnoticed in this study and could only be detected through visual observations (to avoid confusion, we do not present these ranges for single components in Figure~\ref{tree}). As is generally the case in the study of multiple stars, it is only possible to obtain definitive conclusions about the multiplicity order and the structure of the studied systems by complementing different observation techniques. 

In some cases, the companions in intermediate hierarchical levels could affect the dynamical evolution of the inner subsystem through the mechanism known as ``Kozai cycles'' \citep{1962AJ.....67..591K,1962P&SS....9..719L,1997AJ....113.1915I,2000ApJ...535..385F}. In a hierarchical system with non-coplanar outer and inner orbits, the gravitational perturbation on the inner binary produced by the third body leads to the precession of both orbits. If the relative inclination between the orbital planes is in a given critical range, the eccentricity and inclination of the inner binary oscillate, while the period and semimajor axis are not affected. 
However, if the period ratio of the outer and inner orbit is too high, this effect is suppressed by general relativity effects or quadrupolar distortion of the stars. In particular, if the outer period exceeds a minimum value given by $P_\mathrm{out}(\mathrm{yr}) \approx [P_\mathrm{in}(\mathrm{days})]^{1.4}$ Kozai cycles  are not expected to occur \citep{2006Ap&SS.304...75E,2009ApJ...703.1760M}. This condition does not depend significantly on the stellar masses and orbital inclination. Since the eccentricity of the inner binary could achieve very high values as consequence of Kozai cycles \citep{2006Ap&SS.304...75E} 
and considering that the semimajor axis remains constant, the separation of the components at periastron can be significantly reduced. 
As a result, tidal friction  dissipates  energy from the orbital motion leading to a decrease of the semimajor axis and period. Eventually, the orbit will shrink enough for the tidal friction at periastron to become the dominant effect, circularizing the orbit and suppressing the Kozai cycling.

Taking into account the upper limit for $P_\mathrm{out}$ for the existence of Kozai cycles, for each system we determined which is the maximum period that could have a companion in an intermediate hierarchical level to affect the dynamical evolution of the short-period binary subsystem. Particularly, we analysed if this range of possible periods has been covered for our study. We concluded that we can rule out the existence of a companion that is perturbing the inner binary for the subsystems Ca in WDS~15185--4753 and Aab in WDS~20118--6337. For the subsystem Bab of WDS~10209--5603 we considered a tentative period of 3.7 days and we obtained that any companion in an intermediate orbit that could be affecting the dynamical evolution of the inner subsystem should be detected in this study. 
Probably this is the case of the orbit detected through the movement of the barycentre of Bab, so it is an interesting issue to analyse in the future, when the observations of this close subsystem are enough to carry out an orbit determination.

\section{Conclusions}

We have presented the results of a spectroscopic analysis of six systems catalogued as triples in the MSC for which we have detected new components. We found that three components classified as single stars are double-lined spectroscopic binaries, and for one of them we calculated a definitive orbit. 
We obtained RVs curves of both components of a catalogued SB2 subsystem without previous data of the secondary component, and derived new orbital parameters. On the other hand, we detected that other three components are single-lined spectroscopic binaries, for which we obtained preliminary orbits. Additionally, we found one RV variable. Overall, this spectroscopic study has led to the discovery of seven confirmed new components and two members that we expect to confirm with new observations.

This study have revealed that five of the analysed systems are quadruples. 
In one of them, WDS~08314--3904, the membership of one catalogued component was ruled out,  
however, other two additional components have been detected. 
On the other hand, the system WDS~10209--5603 would be formed by five components, distributed 
in four hierarchical levels.

These new results reaffirm the conclusion reached in Paper I about the importance of the analysis of spectroscopic observations over a long time baseline for detecting companions in intermediate hierarchical level that might influence the dynamical evolution of close binary subsystems. We can also see in several of the systems studied here that these ranges of intermediate periods are essentially inaccessible to visual observations. 

This study is a small but interesting evidence that a detailed analysis of the components of multiple systems with early-type components reveals a greater multiplicity degree and a more complex hierarchical structure than known, suggesting that high-order multiple systems are significantly more frequent than it is currently estimated. In particular, the multiplicity order of the analysed systems obtained in this study should be regarded as a lower limit. 
In fact, in all six analysed systems there exists a separation range around $P\sim 10^5$ days that remains virtually unexplored. 
A complete knowledge of these multiples can only be obtained by complementing the spectroscopic results with high-angular resolution visual observations.

\begin{acknowledgements}
This paper was partially supported by a grant from FONCyT-UNSJ PICTO-2009-0125. We are grateful to Natalia Nu\~nez and Ana Collado, who kindly took some spectra used in this investigation.
\end{acknowledgements}

\bibliographystyle{aa}    
\bibliography{biblio1}   

\end{document}